# The Software-Defined Metasurfaces Concept and Electromagnetic Aspects

Anna C. Tasolamprou[1], Alexandros Pitilakis[1,2], Odysseas Tsilipakos[1], Christos Liaskos[1], Ageliki Tsiolaridou[1], Fu Liu[3], Xuchen Wang[3], Mohammad S. Mirmoosa[3], Kypros Kossifos[4], Julius Georgiou[4], Andreas Pitsilides[5], Nikolaos V. Kantartzis[1,2], Dionysios Manessis[6], Sotiris Ioannidis[1], George Kenanakis[1], George Deligeorgis[1], Eleftherios N. Economou[1], Sergei A. Tretyakov[3], Costas M. Soukoulis[1,7] and Maria Kafesaki[1,8]

[1] Foundation for Research and Technology Hellas, 71110, Heraklion, Crete, Greece
[2] Department of Electrical and Computer Engineering, Aristotle University of Thessaloniki, Thessaloniki, Greece
[3] Department of Electronics and Nanoengineering, Aalto University, P.O. Box 15500, Espoo, Finland
[4] Department of Electrical and Computer Engineering, University of Cyprus, 20537, Nicosia, Cyprus
[5] Department of Computer Science, University of Cyprus, 20537, Nicosia, Cyprus
[6] Fraunhofer IZM Berlin, Berlin Center of Advanced Packaging, Gustav-Meyer-Allee 25, Berlin, Germany
[7] Ames Laboratory and Department of Physics and Astronomy, Iowa State University, Ames, Iowa 50011, USA
[8] Department of Materials Science and Technology, University of Crete, 71003, Heraklion, Crete, Greece
*corresponding author, E-mail: atasolam@iesl.forth.gr

## Abstract

We present the concept and electromagnetic aspects of HyperSurFaces (HSFs), artificial, ultrathin structures with software controlled electromagnetic properties. The HSFs key unit is the metasurface, a plane with designed subwavelength features whose electromagnetic response can be tuned via voltage-controlled continuously-tunable electrical elements that provide local control of the surface impedance and advanced functionalities, such as tunable perfect absorption or wavefront manipulation. A nanonetwork of controllers enables software defined HSFs control related to the emerging Internet of Things paradigm.

## 1. Introduction

Metasurfaces are electromagnetically ultrathin structures with designed, subwavelength periodic features that enable exotic functionalities. The constituent parts of the metasurfaces (MS) can be metallic (commonly), dielectric, bulk semiconducting, 2D materials, or a combination of the above, forming subwavelength composite layers [1,2]. Tunable metasurfaces can be realized with the modification of the MS physical features which leads to novel and adjustable functions. Global control of the unit cells can, for example, force tunable perfect absorption function while local control can provide more advanced functionalities such as wavefront manipulation, steering or focusing [3-8]. The tuning mechanisms in a MS are variable and target, in most cases, the material or geometrical parameters that define the MS electromagnetic properties. Some widely known tuning schemes involve, for example, liquid crystals, phase-change materials, MEMS, etc. An effective control mechanism is that of voltage-controlled continuously-tunable electrical elements that modify locally the characteristics of the MS and enable external local control of its resonant features, that is the surface impedance, and therefore of their functionalities [9]. Moreover, the control of the elements can be software driven and the behavior of the metasurface can be defined programmatically; this is the HFS concept. An electronic components nanonetwork in an integrated platform enables the broad vision of smart devices in the emerging Internet of Things paradigm [10]. In this work we present the concept of the Hypersurfaces with a focus on their electromagnetic aspects.

## 2. Discussion

The functional and physical architecture of the Hypersurface tile is presented in Figure 1 and it consists of the metasurface layer, the intratile control layer and the tile gateway layer. We focus on the metasurface layer which realizes the electromagnetic functionalities. Within the concept of the HFSs we examine two different implementations, one is the switch fabric design for operation in the microwave regime and the other is the graphene-based approach which exploits the unique electrically controlled tunable properties of the material's conductivity. We focus mainly on the switch fabric approach since the voltage-controlled continuously-tunable electrical networks are readily implemented in the microwave regime providing the desired local control (contrary to the graphene-based metasurface where local control is particularly challenging). The unit cell of the switch fabric is presented in Figure 1(b). It consists of ultrathin copper patches periodically arranged in a low-loss dielectric substrate which is back-plated by a metallic substrate. When a plane wave impinges on the metasurface, it induces local currents in the patches and the metallic substrate that act as secondary electromagnetic sources modifying the scattered field which leads to the desired operation. The geometrical features of the metasurface are in the order of millimeters and the frequency of operation is 5 GHz. The response of the metasurface to the impinging plane wave is macroscopically described by the complex-valued surface impedance, characterized by real and imaginary parts. Consequently, for the modification of the metasurface response, one needs to adjust dynamically the complex-valued surface impedance. The most efficient way to exert control over both the reactance and resistance is to include two variable lumped elements in the meta-atom, *R* (varistor) and *C* (varactor), which are implemented by controller chips placed behind the backplate. Placing the controller behind the backplate minimizes interference with the impinging electromagnetic wave. By controlling the chip *RC* values within the specification of the electronic circuits we can demonstrate an angle-tunable absorber and advanced functions as anomalous reflection.

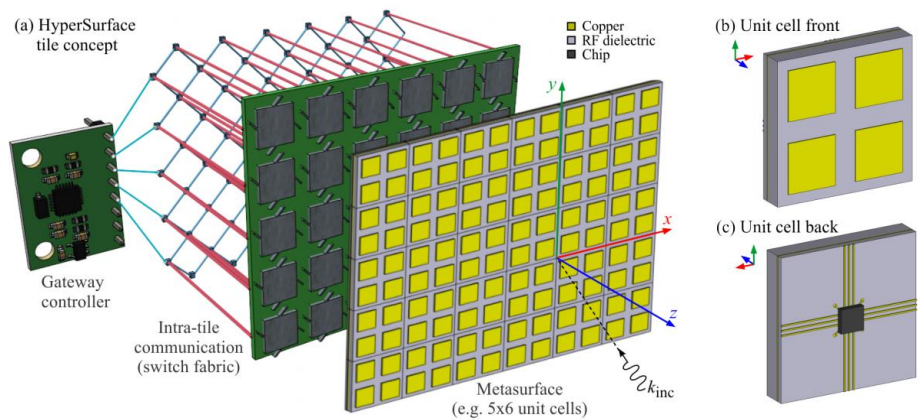

Figure 1: (a) The HyperSurface tile. The switch state configuration setup provides the desired function. A controller intra-network communicates the relevant commands and the inter-tile and external communication is handled by standard gateway hardware. Unit cell of the metasurface (b) front with periodic patches and (c) back with controller chip.

### 3. Conclusions

We presented the concept of Hypersurfaces, software driven metasurfaces with surface impedance that can be modified at will with a set of programmable commands. The control is enabled though a network voltage driven continuously-tunable electrical elements. The electric elements provide variable resistor and capacitor values that modify the complex surface impedance of the metasurface in a local or a global manner. The modification of the surface impedance leads to the adjustment of the metasurface functionalities.

### Acknowledgements

This work was supported by European Union's Horizon 2020 Future Emerging Technologies call (FETOPEN-RIA) under grant agreement no. 736876 (project VISORSURF).